\documentclass[12pt]{article}
\usepackage{amsmath,amssymb,amsfonts}

\newtheorem{theorem}{Theorem}

\newtheorem{example}{Example}

\newcommand{\cA}{\mathcal A}
\newcommand{\cN}{\mathcal N}
\newcommand{\cM}{\mathcal M}
\newcommand{\RR}{\mathbb R}
\newcommand{\ch}{\operatorname{cosh}}
\newcommand{\const}{\operatorname{const}}
\newcommand{\ol}{\overline}

\begin{document}

\title{On a General Theorem of Number Theory
Leading to the Gibbs, Bose--Einstein,
and Pareto Distributions
as well as to the Zipf--Mandelbrot Law
for the Stock Market}

\author{V.~P.~Maslov\thanks{Moscow Institute of Electronics and
Mathematics, pm@miem.edu.ru}} 
\date{}

\maketitle

\begin{abstract}
 The notion of density of a finite set is introduced.
 We prove
a general theorem of set theory which refines the Gibbs,
 Bose--Einstein,
and Pareto distributions as well as the Zipf law.
\end{abstract}

 Suppose that~$\cM^{(n)}$
is a sequence of finite sets tending
as
$n\to  \infty$
to an infinite set.
 Suppose that
$\cN(\cM^{(n)})$
is the number of elements in the set~$\cM^{(n)}$.

 The set~$\cM^{(n)}$
is said to be
$\rho(\,\cdot\,)$-{\it
measurable\/} if there exists a smooth convex function
$\rho(\,\cdot\,)$,
called a {\it density function}, such that the limit
$$
\varliminf_{n\to\infty}\frac{\rho(\cN(\cM^{(n)}))}{\rho(n)}
$$
is finite.
 This limit is called the
$\rho(\,\cdot\,)$-{\it density\/}
of the sequence~$\cM^{(n)}$
of sets.

 Let us present a few examples.

\begin{example}
 Consider the eigenvalues of the
$k$-dimensional oscillator with
potential
$$ 
U(x)=\sum_{i=1}^k(\omega_ix_i^2),\qquad
x\in \RR^k,
$$ 
where the
$\omega_i$
are commensurable:
$$
-\Delta\Psi_i+U(x)\Psi_i=\lambda_i\Psi_i,
\qquad \Psi_i(x)\in L_2(\RR^k).
$$
 Suppose that
$N_\lambda(\lambda_i)$
is the number of its eigenvalues
not exceeding a given positive number~$\lambda$.
 If
$\lambda\to  \infty$,
then the limit
$$
\lim_{\lambda\to\infty}
\frac{\ln N_\lambda(\lambda_i)}{\ln\lambda}=k
$$
coincides with the dimension of the oscillator.
\end{example}

\begin{example}
 Suppose that~$F$
is a compact set and
$N_F(\epsilon)$
is
the minimal number of sets of diameter at most~$\epsilon$
needed to cover~$F$.
 Then
$N_F(\epsilon)$
is
$\ln(\,\cdot\,)$-measurable
and its density coincides with the metric order of the compact set~$F$
(see~\cite{1}).
\end{example}

 Consider the set
$\{\cM_1\}$
of nonnegative numbers
$\lambda_1,\lambda_2,\dots,\lambda_n$
and the set
$\{\cM_2\}$
of integers
$1,2,\dots,N$;
\
$N=N(n)$.
 Suppose that the set
$\{\cM_2\}$
is
$\ln(\,\cdot\,)$@-commensurable
and~$s$
is its density:
$$
\varliminf_{n\to\infty}\frac{\ln N}{\ln n}=s.
$$
 Besides, let
$\ol\lambda(n)$
be the arithmetic mean of the ensemble of~$\lambda_i$:
$$
\ol\lambda(n)=\frac1n\sum_1^n\lambda_i.
$$

 Suppose that we are given a number
$E(n)$.
 Consider the following
cases:
\begin{description}
\item{(1)}
$\epsilon\le E(n)\le\ol\lambda(n)N$,
\
$\epsilon>  0$;
\item{(2)}
$E(n)\ge\ol\lambda(n)N$.
\end{description}

 Consider the set of mappings of
$\{\cM_2\}$
onto
$\{\cM_1\}$.
 Two mappings are said to be {\it equivalent\/} if their images
are identical.
 Further, we shall only consider nonequivalent mappings
and denote them by
$\{\cM_3\}$.

 Suppose that the sum of elements in
$\{\cM_2\}$
is equal to
$N=\sum_{i=1}^nN_i$
and the bilinear form of the pair of sets
$\{\cM_1\}$
and
$\{\cM_2\}$
satisfies the condition
$$
\biggl|\sum_{i=1}^nN_i\lambda_i\biggr|\le E(n)
$$
in case~(1) and
$$
\biggl|\sum_{i=1}^nN_i\lambda_i\biggr|\ge E(n)
$$
in case~(2).

 Note that the set
$\{\cM_3\}$
is
$\ln\ln(\,\cdot\,)$-measurable.

 Without loss of generality, we assume that the real numbers
$\lambda_1,\lambda_2,\dots,\lambda_n$
are naturally ordered, i.e.,
$0\le\lambda_i\le  \lambda_{i+1}$,
and we split the interval
$1,2,\dots,n$
into~$k$
intervals
(to within~1), where
$k$
is independent of
$n$:
\begin{gather*}
1,2,\dots,n_1, \quad
n_1+1,n_1+2\dots,n_2, \quad
n_2+1,n_2+2,\dots,n_3, \quad \dots,
\\
n_{k-1}+1,n_{k-1}+2,\dots,n_k,
\qquad
\sum_{l=1}^kn_l=n;
\end{gather*}
here
$l=1,\dots,k$
is the number of the interval.

 Denote by~$\ol\lambda_l$,
$l=1,2,\dots,k$,
the nonlinear average
of~$\lambda_i$
over each interval:
$$
\ol\lambda_l
=\Phi_{\alpha\beta}\biggl(\sum^{n_l}_{n_{l-1}}
\psi_{\alpha\beta}(\lambda_i)\biggr),
$$
where
$\psi_{\alpha\beta}(x)$
is the two-parameter family of functions
and~$\Phi_{\alpha\beta}$
is its inverse:
$\Phi_{\alpha\beta}(\psi_{\alpha\beta}(x))=  1$;
namely,
\begin{align}
\text{(a)} \quad \psi_{\alpha\beta}
&=\alpha e^{-\beta x} &\qquad& \text{for}\quad s>1;
\tag{1}
\\
\text{(b)} \quad \psi_{\alpha\beta}
&=\frac1{\alpha e^{\beta x}-1} &\qquad& \text{for}\quad s=1;
\notag
\\
\text{(c)} \quad \psi_{\alpha\beta}
&=\frac1{\beta x+\ln\alpha} &\qquad& \text{for}\quad 0<s<1.
\notag
\end{align}
 The parameters~$\alpha$
and~$\beta$
are related to
$N(n)$
and
$E(n)$
by the conditions
\begin{equation}
\sum^n_{i=1}\psi_{\alpha\beta}(\lambda_i)=N(n), \qquad
\sum^n_{i=1}\lambda_i\psi_{\alpha\beta}(\lambda_i)=E(n).
\tag{2}
\end{equation}
 Consider the subset
$\cA\subset  \cM_3$:
\begin{equation}
\cA=\biggl\{\sum_{l=1}^k\biggl(\sum^{n_l}_{n_{l-1}}N_i
-\psi_{\alpha\beta}(\ol\lambda_l)\biggr)\le\Delta\biggr\},
\tag{3}
\end{equation}
where
\begin{equation}
\Delta=\begin{cases}
\sqrt N\ln^{1/2+\epsilon}N& \text{for}\; \;N\ll n,\\
\sqrt n\ln^{1/2+\epsilon}n& \text{for}\; \;N\sim n,\\
\dfrac N{\sqrt n}\ln^{1/2+\epsilon}n& \text{for}\; \;N\gg n
\end{cases}
\tag{4}
\end{equation}
is called the {\it resolving power}.

\begin{theorem}
 The following inequality holds{\rm:}
$$
\frac{\cN(\cM_3\setminus\cA)}{\cN(\cM_3)}
\le\frac C{n^k}+\frac C{N^k},
$$
where $k$ is arbitrary and
$C$
is a constant independent of~$n$ and~$N$.
\end{theorem}

{\bf Proof.}
 Obviously,
\begin{align}
\cN\{\cM_3\setminus A\}
&=\sum_{\{N_i\}}\biggl(\Theta\biggl\{N(n)E(n)
-\sum_{i=1}^nN_i\lambda_i\biggr\}
\delta_{\sum_{i=1}^nN_i,N(n)}\tag{5}
\\ &\qquad
{\sum_{\{N_i\}}\biggl(}
\times
\Theta\biggl\{\biggr|\sum_{l=1}^k\biggl(\sum_{i=n_{l-1}}^{n_l}N_i
-\psi_{\alpha\beta}(\ol\lambda_l)\biggr)\biggr|
-\Delta\biggr\}\biggr).
\notag
\end{align}
 Here the sum is taken over all integers~$N_i$,
$\Theta(x)$
is the Heaviside function,
$$
\Theta(x)=\begin{cases}
1& \text{for}\; \;x\ge0\\
0& \text{for}\; \;x<0,
\end{cases}
$$
and
$\delta_{k_1,k_2}$
is the Kronecker delta,
$$
\delta_{k_1,k_2}=\begin{cases}
1& \text{for}\; \;k_1=k_2,\\
0& \text{for}\; \;k_1\ne k_2.
\end{cases}
$$
 Let us use the integral representations
\begin{gather}
\delta_{N,N'}
=\frac{e^{-\nu N}}{2\pi}\int_{-\pi}^\pi d\phi\,
e^{-iN\phi}e^{\nu N'}e^{iN'\phi},
\tag6
\\
\Theta(y)
=\frac1{2\pi i}\int_{-\infty}^\infty dx\,
\frac1{x-i}e^{\beta y(1+ix)}.
\tag7
\end{gather}
 We have
\begin{equation}
\int_0^\infty dE\,
\Theta\biggl(E-\sum_{i=1}^nN_i\lambda_i\biggr)e^{-\beta E}
=\int_{\sum_{i=1}^nN_i\lambda_i}^\infty dE\,e^{-\beta E}
=\frac{e^{-\beta\sum_{i=1}^nN_i\lambda_i}}\beta.
\tag{8}
\end{equation}

 Denote
\begin{gather*}
Z(\beta,N)=\sum_{\{N_i\}}e^{-\beta\sum_{i=1}^nN_i\lambda_i},
\qquad
\zeta(\nu,\beta)=\prod^k_{l=1}\zeta_l(\nu,\beta),
\\
\zeta_l(\nu,\beta)
=\prod_{i=n_{l-1}}^{n_l}\xi_i(\nu,\beta),
\qquad
\xi_i(\nu,\beta)=\frac1{1-e^{\nu-\beta\lambda_i}},
\quad i=1,\dots,n,
\end{gather*}
and
$$
\Gamma(E,N)=\cN\{\cM_3\}.
$$
 Since
$\cN\{\cM_3\}(E)<\cN\{\cM_3\}(E+  \epsilon)$
for
$\epsilon>  0$,
we have
\begin{equation}
Z(\beta,N)
\ge\beta\int_E^\infty dE'\,\Gamma(E',N)e^{-\beta E'}
=\Gamma(E,N)e^{-\beta E}.
\tag{9}
\end{equation}
 Therefore,
\begin{equation}
\cN\{\cM_3\}\le Z(\beta,N)e^{\beta E}.
\tag{10}
\end{equation}
 But, by~\thetag{6},
\begin{equation}
Z(\beta,N)
=\frac{e^{-\nu N}}{2\pi}\int_{-\pi}^\pi d\alpha\,
e^{-iN\alpha}\zeta(\beta,\nu+i\alpha);
\tag{11}
\end{equation}
hence
\begin{align}
&
\cN\{\cM_3\setminus\cA\}
\tag{12}\\ 
&\quad
\le\biggl|\frac{e^{-\nu N+\beta E}}{2\pi}
\int_{-\pi}^\pi\biggl[\exp(-iN\phi)
\sum_{\{N_j\}}\biggl(\exp\biggl\{(-\beta\sum_{j=1}^n
N_j\lambda_j)+(i\phi+\nu)N_j\biggr\}\biggr]\,d\phi
\notag
\\ &\qquad \quad\times
\Theta\biggl\{\biggl|\sum_{l=1}^k
\biggl(\sum_{j=n_{l-1}}^{n_l}N_j
-\psi_{\alpha\beta}(\ol\lambda_l)\biggr)\biggr|
-\Delta\biggr\}\biggr)\biggr|,
\notag
\end{align}
where
$\beta$
and
$\nu=-\ln\alpha$
are real parameters for which
the series is convergent.

 Estimating the right-hand side, carrying the modulus through
the integral sign and then through the sign of the sum, and
integrating
over~$\phi$,
we obtain
\begin{align}
\cN\{\cM_3^n\setminus\cA\}
&\le e^{-\nu N}\exp\beta E(n)
\sum_{\{N_i\}}\exp\biggl\{-\beta\sum_{i=1}^nN_i\lambda_i
+\nu N_i\biggr\}
\tag{13}
\\ &\qquad\times
\Theta\biggl\{\biggl|\sum_{l=1}^k\biggl(\sum_{i=n_{l-1}}^{n_l}N_i
-\psi_{\alpha\beta}(\ol\lambda_l)\biggr)\biggr|-\Delta\biggr\}.
\notag
\end{align}
 Let us use the following inequality for the hyperbolic cosine
$\ch(x)=(e^x+e^{-x})/2$:
\begin{equation}
\prod_{l=1}^k\ch(x_l)\ge2^{-k}e^\delta
\qquad \text{for all}\quad x_l,
\quad \sum_{l=1}^k|x_l|\ge\delta\ge0.
\tag{14}
\end{equation}
 Hence, for all positive~$c$
and~$\Delta$,
we have the
inequality (cf\.~\cite{2,3})
\begin{equation}
\Theta\biggl\{\biggl|\sum_{l=1}^k\biggl(\sum_{i=n_{l-1}}^{n_l}N_i
-\psi_{\alpha\beta}(\ol\lambda_l)\biggr)\biggr|-\Delta\biggr\}
\le2^ke^{-c\Delta}\prod_{l=1}^k\ch\biggl(c\sum_{i=n_{l-1}}^{n_l}N_i
-c\psi_{\alpha\beta}(\ol\lambda_l)\biggr).
\tag{15}
\end{equation}
 We obtain
\begin{align}
\cN\{\cM_3\setminus\cA\}
&\le2^ke^{-c\Delta}\exp(\beta E(n)-\nu N)
\tag{16}
\\ 
&\qquad\times
\sum_{\{N_i\}}\exp\biggl\{-\beta\sum_{i=1}^nN_i\lambda_i
+\nu N_i\biggr\}
\prod_{l=1}^k\ch\biggl(\sum_{i=n_{l-1}}^{n_l}cN_i
-c\psi_{\alpha\beta}(\ol\lambda_l)\biggr)
\notag
\\ 
& =e^{\beta E(n)}e^{-\nu N}e^{-c\Delta}
\notag
\\ 
\vspace{-2.5pt} &\qquad\times
\prod_{l=1}^k\bigl(\zeta_l(\nu+c,\beta)
\exp(-c\psi_{\alpha\beta}(\ol\lambda_l))
+\zeta_l(\nu-c,\beta)
\exp(c\psi_{\alpha\beta}(\ol\lambda_l))\bigr).
\notag
\end{align}
 Let us apply Taylor's formula to
$\zeta_l(\nu+c,\beta)$.
 Namely,
there exists a
$\gamma<  1$
such that
$$
\ln(\zeta_l(\nu+c,\beta))
=\ln\zeta_l(\nu,\beta)
+c(\ln\zeta_l)'_\nu(\nu,\beta)
+\frac{c^2}2(\ln\zeta_l)^{''}_\nu(\nu+\gamma c,\beta).
$$
 Obviously,
$$
\frac\partial{\partial\nu}\ln\zeta_l
\equiv\psi_{\alpha,\beta}(\ol\lambda_l).
$$
 Let
$c=\Delta/D(\nu,\beta)$,
where
$D(\nu,\beta)=(\ln\zeta)_\nu^{''}(\nu,\beta)$.
 The right-hand side of relation~\thetag{16} is equal to
$$
2^ke^{\beta E(n)}e^{-\nu N}\prod_{l=1}^k
\zeta_l(\nu,\beta)\exp\biggl\{-\frac{\Delta^2}{D(\nu,\beta)}
+\frac{\Delta^2D(\nu+\gamma\Delta/D(\nu,\beta),\beta)}
{2(D(\nu,\beta))^2}\biggr\}.
$$
 Imposing
the following constraint on~$\Delta$:
$$
D\biggl(\nu+\frac\Delta{D(\nu,\beta)},\beta\biggr)
\le(2-\epsilon)D(\nu,\beta),
$$
where
$\epsilon>  0$,
and taking into account the
fact that
$D(\nu,\beta)$
is monotone increasing in~$\nu$,
we finally obtain
$$
\cN(\cM_3\setminus\cA)
\le2^ke^{\beta E(n)}e^{-\nu N}\zeta(\nu,\beta)
e^{-\epsilon\Delta^2/D(\nu,\beta)}.
$$
 Next, let us estimate
$\zeta(\nu,\beta)$.

 The following lower bound for
$Z(\beta,N)$
was obtained in~\cite{2}, relation~(95):
\begin{equation}
\zeta(\nu',\beta)
\le\sqrt{27D(\nu',\beta)}Z(\beta,N)e^{\nu'N},
\tag{17}
\end{equation}
where
$\nu'=\nu'(\beta,N)$
is determined from the condition
\begin{equation}
\sum_{i=1}^n\xi_i(\nu,\beta)=N.
\tag{18}
\end{equation}
 Suppose that
$\beta=  \beta'$
is determined from
the condition
\begin{equation}
\sum_{i=1}^n\lambda_i\xi_i(\nu,\beta)=E(n).
\tag{19}
\end{equation}

 Since
$Z(\beta,N)$
is determined by the integral~\thetag{11},
its asymptotics given by the saddle-point method
(the stationary phase due to Laplace) yields a unique saddle point for
$\alpha=  0$.

 The square root of the second derivatives with respect to~$\alpha$
will appear in the denominator.
 As a result, we obtain
$$
\zeta(\nu',\beta')
\le Ce^{N\nu'}e^{-\beta'E(\mu)}D(\nu',\beta')\cN\{\cM_3\},
$$
where
$C$
is a constant.
 Therefore, we finally obtain
\footnote{A lower bound
for
$\cN\{\cM_3\}$
was obtained by G.~V.~Koval' and the author
in~\cite{4} without recourse to the saddle-point method.}
\begin{equation}
\frac{\cN\{\cM_3\setminus\cA\}}{\cN\{\cM_3\}}
\le2^kCD(\nu',\beta')
e^{-\epsilon\Delta^2/D(\nu',\beta')}.
\tag{20}
\end{equation}
 Further, it is easy to estimate
$D(\nu,\beta)$
as a function of
$N,n$:
$D\sim  N$
for
$N<  n$,
while,
for
$n\gg  N$,
the estimate for~$D$
yields the
relation
$D\sim N^2/n$.
 Hence we obtain the estimate for
$\cN(\cM_3\setminus\cA)/\cN\{\cM_3\}$,
given in the theorem.

\begin{example}
 For the case in which
$s>  0$
is sufficiently small
(and hence,
$\sum_{l=1}^kN_l=  N$
not very large),
the Bose--Einstein distribution is of the form
\begin{equation}
\frac{N_l}{N_{l+1}}
\sim\frac{\sum_{i=n_{l-1}}^{n_l}e^{-\lambda_i\beta}}
{\sum_{i=n_{l-1}}^{n_l}e^{-\lambda_i\beta}},
\tag{21}
\end{equation}
where
$\beta=1/(kT)$,
$T$
is the temperature, and
$k$
is the Boltzmann
constant.

 In the case of a Bose gas, for
$s<  1$,
we have a
distribution of Gibbs type, i.e., the ratio of the number of particles
on the~$l$th interval to the number of particles on
the
$(l+  1)$th interval obeys formula~\thetag{21}.
\end{example}

\begin{example}
 In the case
$s>  1$,
we obtain a refinement of the
Zipf--Mandelbrot law~\cite{5}, namely,
\begin{equation}
\frac{N_l}n
\sim\biggl\{\sum_{i=n_{l-1}}^{n_l}
\frac1{\lambda_i+\nu}\biggr\}.
\tag{22}
\end{equation}
 However, if
$s$
is close to~1, then it is better to use relation~(b)
in~\thetag{1}, which uniformly passes into relation~(c) and
relation~(a).

 Note that if all the~$\lambda_i$
on the
$l$th interval
are identical and equal to~$\lambda^{(l)}$,
then
$N_l/n_l\sim1/\lambda^{(l)}$,
and since
$n_l\sim  N_l^{1/s}$,
it follows that, in
this case, we obtain the Zipf--Mandelbrot formula.
\end{example}

\begin{example}{(relation
between the sales volume and the prices on the stock market)}
 Let us now consider the relation between the prices and the number
of sold (bought) shares of some particular company on the stock
market.

 Since the number~$n_i$
of sold shares of that company during the
$i$th day is equal to the number of bought shares and~$\lambda_i$
is the price of the shares at the end of the day, we set, averaging
over
$n_l$~days, the nonlinear average price as
$$
\ol\lambda^l=\Phi\biggl(\sum_{i=n_l}^{n_{l+1}}\phi(\lambda_i)\biggr),
$$
where
$\phi(x)=1/(x+  \nu)$,
$\nu=  \const$,
and
$\Phi(x)$
is the function inverse to
$\phi(x)$.
 Then, by
Theorem~1, we have
\begin{equation}
n_l\simeq A\phi(\ol\lambda_i^l),
\tag{23}
\end{equation}
where
$A$
is a constant.

 Thus, the stock market obeys the refined Zipf--Mandelbrot law
if all the types of transactions are equiprobable (see~\cite{6,7}).
\end{example}

 In conclusion, note that although the theorem is stated in terms
of set theory owing to the fact that we have introduced the notion
of equivalent mappings, it belongs, most likely, to number theory.
 Under the same conditions, considering the set of mappings of the
set~$\cM_1^{(n)}$
onto the set~$\cM_2^{(n)}$
without the condition for the equivalence
of mappings, i.e., considering all mappings, we can obtain a similar
theorem that will only be relevant to the refinement of the Gibbs
distribution.
 At the same time, such a theorem is related to
information theory and a generalization of Shannon's entropy.
 Here the estimate has special features, and the corresponding article
will be
published jointly with G.~V.~Koval'.


\begin{thebibliography}{99}


\bibitem{1}
W.~Hurewicz and H.~Wallman, {\it Dimension Theory}, Princeton
Mathematical Series, vol. 4. Princeton University Press, Princeton,
NJ, 1941; Russian translation: Moscow, 1948.

\bibitem{2}
V.~P.~Maslov, ``Nonlinear averages in economics,'' {\it Mat.
Zametki\/}
[{\it Math. Notes\/}], {\bf 78} (2005), no.~3, 377--395.

\bibitem{3}
V.~P.~Maslov, ``The law of large deviations in number theory.
Computable functions of several arguments and decoding,'' {\it
Dokl. Ross. Akad. Nauk\/} [{\it Russian Acad. Sci. Dokl. Math.}], {\bf
404} (2005), no.~6, 731--736.

\bibitem{4}
G.~V.~Koval' and V.~P.~Maslov, ``On estimates for a large
partition function,'' (to appear).

\bibitem{5}
B.~Mandelbrot, {\it Structure formelle des textes et communication},
Word, vol.~10. no.~1, New York, 1954.

\bibitem{6}
V.~P.~Maslov, ``The principle of increasing complexity of portfolio
formation on the stock exchange,'' {\it Dokl. Ross. Akad. Nauk\/}
[{\it Russian Acad. Sci. Dokl. Math.}], {\bf 404} (2005), no.~4,
446--450.

\bibitem{7}
V.~P.~Maslov, ``A refinement of the Zipf law for frequency
dictionaries
and stock exchanges,'' {\it Dokl. Ross. Akad. Nauk\/} [{\it Russian
Acad. Sci. Dokl. Math.}], {\bf 405} (2005), no.~5.

\end{thebibliography}
\end{document}